\documentclass[a4,12pt]{article}
\usepackage{amsmath,amssymb,color,graphics,amscd,amsfonts,epsf,indentfirst}
\usepackage{epsfig}
\usepackage[dvipdfm,hypertex]{hyperref}
\allowdisplaybreaks[4]

\makeatletter
\@addtoreset{equation}{section}

\makeatletter
\renewcommand\section{\@startsection {section}{1}{\z@}%
                                   {-3.5ex \@plus -1ex \@minus -.2ex}
                                   {2.3ex \@plus.2ex}%
                                   {\normalfont\large\bfseries}}
\renewcommand\subsection{\@startsection{subsection}{2}{\z@}%
                                     {-3.25ex\@plus -1ex \@minus -.2ex}%
                                     {1.5ex \@plus .2ex}%
                                    {\normalfont\bfseries}}

\def\no{\nonumber \\}
\def\btab{\begin{table}[h] \begin{center} \begin{tabular}{l lp{3in}}}
      \def\etab{\end{tabular} \end{center} \end{table}}
\def\btabm{\begin{center} \begin{tabular}}
    \def\etabm{\end{tabular} \end{center}}

\def\ie{{\it i.e.\ }}

\def\a{{\alpha}}
\def\al{\alpha'}
\def\b{{\beta}}
\def\g{{\gamma}}
\def\G{{\Gamma}}

\def\D{{\Delta}}
\def\ep{{\epsilon}}

\def\m{{\mu}}
\def\n{{\nu}}

\def\ph{{\phi }}

\def\t{{\theta}}

\def\om{\omega}

\def\vp{\varphi}

\def\f#1#2{{\frac{#1}{#2}}}

\def\s{\sqrt}

\def\f {\frac}
\def\ti{\tilde}

\def\p{\partial}
\def\we{\wedge}

\def\CM{{\cal M}}
\def\CN{{\cal N}}
\def\ap{\alpha}

\newcommand{\be}{\begin{equation}}
\newcommand{\ba}{\begin{eqnarray}}
\newcommand{\ea}{\end{eqnarray}}
\newcommand{\ee}{\end{equation}}

\begin{document}

\begin{titlepage}
  \thispagestyle{empty}

  \begin{flushright}
    KUNS-2153\\
  \end{flushright}

  \vspace{2cm}

  \begin{center}
    \font\titlerm=cmr10 scaled\magstep4
    \font\titlei=cmmi10 scaled\magstep4
    \font\titleis=cmmi7 scaled\magstep4
    \centerline{\titlerm
      Fuzzy Ring from M2-brane Giant Torus
      \vspace{2.2cm}}
    \noindent{{
        Tatsuma Nishioka\footnote{e-mail:nishioka@gauge.scphys.kyoto-u.ac.jp}
        and Tadashi Takayanagi\footnote{e-mail:takayana@gauge.scphys.kyoto-u.ac.jp}
      }}\\
    \vspace{0.8cm}

    {\it Department of Physics, Kyoto University, Kyoto 606-8502, Japan}

    \vspace{1cm}
    {\large \today}
  \end{center}

  \vskip 5em

  \begin{abstract}
    We construct spinning dual M2 giant gravitons in $AdS_4\times
    S^7$, which generically become $\f{1}{16}$ BPS states,
    and show that their world-volumes become torii. By taking
    an orbifold, we obtain
    spinning dielectric D2-brane configurations in $AdS_4\times CP^3$
    dual to specific BPS operators in ABJM theory.
    This reveals a novel mechanism
    how to give an angular momentum to a dielectric D2-brane.
    We also find that when its
    angular momentum in the $AdS_4$ becomes large, it approaches to
    a ring-like object. Our result might suggest
    an existence of supersymmetric black rings
    in the $AdS_4$ background. We will also discuss dual giant
    gravitons in $AdS_4\times CP^3$.
  \end{abstract}

\end{titlepage}




\section{Introduction}

The studies of D-branes and M-branes have certainly been crucial in
recent developments of string theory and M-theory. The classical
dynamics of a brane is described by its world-volume theory, which
is reparametrization invariant. At the same time, they are the
sources of various gauge fields such as RR-fields. Thus the dynamics
of their world-volumes is largely affected by the presence of the
various fluxes in string theory and M-theory.

In particular, consider a D2-D0 bound state in type IIA string with
the RR-3 form flux in the world-volume direction. Then it is
well-known that the stable world-volume is given by a two sphere,
called the fuzzy sphere (or dielectric D2-brane) \cite{My}. This
configuration can be regarded as a system of multiple D0-branes,
which are expanded into a sphere due to the non-abelian RR coupling
\cite{My}. In the beginning of this paper, we will explicitly
construct a supersymmetric fuzzy sphere in the fully back-reacted
IIA background of $AdS_4\times CP^3$ \cite{NPW,ABJM}. This predicts
a new class of supersymmetric states in the dual ABJM theory
\cite{ABJM}.

 We may naturally think a fuzzy sphere as a rigid body in the
low energy limit since it is composed of infinitesimally small and
massive constituents (\ie D0-branes). This suggests, for example,
that we can give it an angular momentum. Actually, by colliding
closed strings with the fuzzy sphere, we can excite the angular
momentum. However, naively this contradicts with the D2-brane
description because the world-volume theory is reparametrization
invariant and there seems no room for adding any angular momenta.

\begin{figure}[htbp]
   \begin{center}
     \includegraphics[height=6cm]{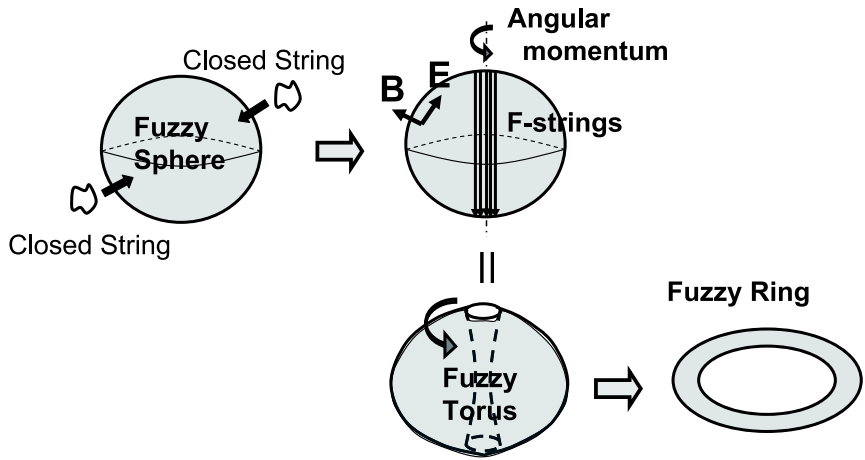}
   \end{center}
   \caption{If we collides closed strings with a fuzzy sphere (or
    a dielectric D2-brane), it should begin
     spinning. To give a non-vanishing angular momentum, we need
     fundamental strings which connect between the North Pole and the
     South Pole. By regarding the total system as a bound state of
     a D2-brane, D0-branes and F-strings, its world-volume becomes
    a torus. In this paper we will present an exact profile of this
     configuration by solving the BPS equation.}\label{fuzzyspin}
 \end{figure}

One of the main purposes of this paper is to resolve this puzzle
completely. We argue that the spinning fuzzy `sphere' is realized by
putting a non-vanishing electric flux so that it produces the
non-zero Poynting vector together with the magnetic flux due to the
D0-brane charge (see Fig.\ref{fuzzyspin}). Even though this is
similar to the mechanism of the supertube \cite{MaTo}, our case is
far more non-trivial since the Gauss law on the sphere seems to
contradict the presence of the electric flux. Indeed, as we will
show explicitly, the topology of the world-volume is no longer a
sphere but it should be changed into a torus in order to realize a
stationary spinning configuration. This topology change is naturally
explained by interpreting the sources of electric flux as the
fundamental strings attached on the North and South Poles of the two
sphere as explained in Fig.\ref{fuzzyspin}. When its angular
momentum in the $AdS_4$ becomes large, the fuzzy torus degenerates
into a ring-like object (`fuzzy ring'). Since some of such fuzzy
rings become BPS states, they may resemble supersymmetric black
rings \cite{BR}.

Another D-brane configuration, which is analogous to the previous
example, will be the giant gravitons \cite{MST,GMT,HHI,MaSu}. Here
we especially consider\footnote{As usual, a dual giant graviton
means a giant graviton whose $2+1$ dimensional world-volume expands
in the $AdS_4$ direction \cite{HHI}.} dual giant gravitons in
$AdS_4\times S^7$. It is a spherical M2-brane with angular momenta
(or R-charges in the dual $CFT_3$) in the $S^7$ direction. As we
will show in this paper, its reduction to the type IIA string via
the orbifolding procedure introduced in \cite{ABJM}, precisely leads
to the mentioned spherical dielectric D2-brane in $AdS_4\times
CP^3$.

 Moreover, we will construct a spinning dual
 giant gravitons in M-theory by solving the BPS equations explicitly.
 By reducing them to type IIA via the orbifold,
 we will obtain the exact solution of the spinning fuzzy D2-brane.
 We will show
 that the world-volumes of a spinning giant graviton becomes\footnote{
The spike and torus configurations of D-branes or M-branes have been
already noticed by studying the giant gravitons in the pp-wave
backgrounds in \cite{MiG,TaTa,SaJa,AlJa,Gpp}. Also we would like to
refer to \cite{Gwave,KiLe,GGKM,AsSu} for excellent constructions of
generic dual giant gravitons in $AdS_5\times S^5$  based on the
approach \cite{Mi} using the holomorphic surface.} a sphere (see
Fig.\ref{GS3d}) with two spikes attached or a torus (see
Fig.\ref{GT3d}).

   This paper is organized as follows: In section two, we briefly review the
   M-theory on $AdS_4\times S^7$ and type IIA string on $AdS_4\times
   CP^3$. In section three, we construct
   a dielectric D2-brane solution in $AdS_4\times CP^3$
   and show that it can be obtained from the reduction of
   a dual giant graviton in $AdS_4\times S^7$ by the orbifolding.
   In section four, we construct spinning dual giant gravitons
   in $AdS_4\times S^7$ by solving the BPS equations.
   In section five, we obtain the spinning dielectric D2-brane solution
   in the $AdS_4\times CP^3$ by
   taking the orbifold reduction of the spinning giant graviton
   solutions. In section 6, we construct dual giant gravitons in
   type IIA string on $AdS_4\times CP^3$.
   In section 7, we summarize the conclusion.

\section{Review of M-theory on $AdS_4\times S^7/Z_k$ and IIA on $AdS_4\times
CP^3$}

The near horizon metric of $N'(=Nk)$ M2-branes becomes $AdS_4 \times
S^7$
\begin{align}\label{metric}
ds^2&=\f{R^2}{4}\left[ds^2_{AdS_4}+4 d\Omega_7^2\right],\qquad
R = l_p(2^5\pi^2N')^{\f{1}{6}},\\
ds^2_{AdS_4} &= -\left(1+ r^2\right) dt^2
+\f{dr^2}{1+r^2}+r^2(d\t^2 + \sin^2\theta d\vp^2),\nonumber
\end{align}
with 4-form flux
\begin{align}
  F^{(4)} = -\f{3R^3}{8}r^2\sin\t dt\we dr \we d\t \we d\vp,
\end{align}
where $\Omega_7$ represents the coordinate of $S^7$ with unit
radius:
\begin{align} |z_1|^2+|z_2|^2+|z_3|^2+|z_4|^2=1. \end{align}
We can express \begin{align} z_i=\mu_i e^{i\xi_i}, \ \ \ \
(i=1,2,3,4) \label{corde}
\end{align} where $(\mu_1,\mu_2,\mu_3,\mu_4)=(\sin\alpha, \cos\alpha
\sin\beta,\cos\alpha\cos\beta\sin\gamma,
\cos\alpha\cos\beta\cos\gamma)$. We choose the coordinate
$(x^0,x^1,...,x^{10})$ as follows
\begin{align}
& x^0=t,\ \ x^1=r, \ \ x^2=\theta,\ \ x^3=\vp,\ \ x^4=\alpha,\ \ \no
& x^5=\beta,\ \ x^6=\gamma, \ \ x^7=\xi_1,\ \ x^8=\xi_2,  \
x^9=\xi_3,\ \ x^{10}=\xi_4. \label{corc}
\end{align}
Our convention of $\Gamma$ matrices is such that $\gamma^\mu$ are
normalized such that $\{\gamma_\mu,\gamma_\nu\}=2\eta_{\mu\nu}$; on
the other hand, $\Gamma^\mu$ are not normalized \ie
$\{\Gamma_\mu,\Gamma_\nu\}=2g_{\mu\nu}$. Then the Killing spinors
$\ep$ on $AdS_4\times S^7$ are found to be (see appendix
\ref{Ap:Killing})
\begin{align}
\ep=e^{\f{\alpha}{2}\hat{\gamma}
\gamma_4}e^{\f{\beta}{2}\hat{\gamma}  \gamma_5}
e^{\f{\gamma}{2}\hat{\gamma} \gamma_6}e^{\f{\xi_1}{2}\gamma_{47}}
e^{\f{\xi_2}{2}\gamma_{58}}e^{\f{\xi_3}{2}\gamma_{69}}
e^{\f{\xi_4}{2}\hat{\gamma} \gamma_{10}}e^{-\f{\rho}{2}\gamma_1
\hat{\gamma}} e^{-\f{t}{2}\gamma_0 \hat{\gamma}}
e^{\f{\theta}{2}\gamma_{12}} e^{\f{\vp}{2}\gamma_{23}}\ep_0
\equiv \CM \ep_0,
\end{align}
where we defined $\hat{\gamma}\equiv\gamma^{0123}$ and $\sinh
\rho=r$; $\ep_0$ is a constant 11D spinor which satisfies
$\g_{012345678910}\ep_0 = \ep_0$. It is easy to check $\{\hat \g,
\g_a\}=0, (a=0,1,2,3)$. As we can choose a constant spinor $\ep_0$
arbitrary, there are 32 Killing spinors in $AdS_4 \times S^7$
as is well-known.

Let us take the $Z_k$ orbifold of $S^7$ and reduce the M-theory
background $AdS_4\times S^7/Z_{k}$ to the type IIA string background
$AdS_4 \times CP^3$ following \cite{ABJM}. The $Z_k$ quotient acts
on the complex coordinates $z_i~(i=1,2,3,4)$ as
\begin{align}
  z_i \sim e^{i\f{2\pi}{k}}z_i,
\end{align}
thus $\xi_i$ is identified as  $\xi_i \sim \xi_i + \f{2\pi}{k}$
under the orbifold action. If we define
$\gamma_{47}\ep_0=is_1\ep_0$, $\gamma_{58}\ep_0=is_2\ep_0$ and
$\gamma_{69}\ep_0=is_3\ep_0$, the spinors which survive the $Z_k$
orbifold projection are the ones
$(s_1,s_2,s_3)=(+,+,-),(+,-,+),(-,+,+),(+,-,-),(-,+,-),(-,-,+)$. The
ones with $(+,+,+)$ and $(-,-,-)$ are projected out. Therefore 24
out of 32 Killing spinors are survived in the $AdS_4\times S^7/Z_k$
with $k>2$ as claimed in \cite{ABJM}.

To see the reduction of our M-theory background to the type IIA
string explicitly, it is easier to parameterize $S^7$ as
follows\footnote{ In this paper, however, we will always work with
the coordinate choice (\ref{corc}) and (\ref{corde}) except in this
section and section 6.} (see \cite{NiTa}) instead of (\ref{corde})
\begin{align}
  z_1&=\cos\zeta~ \cos\f{\theta_1}{2}~e^{i\f{\chi_1+\vp_1}{2}}, \qquad
  z_2=\cos\zeta~ \sin\f{\theta_1}{2}~e^{i\f{\chi_1-\vp_1}{2}}, \no
  z_3&=\sin\zeta~ \cos\f{\theta_2}{2}~e^{i\f{\chi_2+\vp_2}{2}}, \qquad
  z_4=\sin\zeta~ \sin\f{\theta_2}{2}~e^{i\f{\chi_2-\vp_2}{2}},
  \label{angles}
\end{align}
where the angular valuables run the values $0\leq \zeta
<\f{\pi}{2}$, $0\leq \chi_i <4\pi$, $0\leq \vp_i \leq 2\pi$ and
$0\leq \theta_i<\pi$.

Define new coordinates
\begin{align}
  \chi_1=2y+\psi,\ \ \ \ \ \ \ \chi_2=2y-\psi.
\end{align}
The $Z_k$ orbifold action is now given by $y\sim y+\f{2\pi}{k}$.
Then
\begin{align}\label{S7}
  ds_{S^7}^2=ds^2_{CP^3}+(dy+A)^2,
\end{align}
where $A$ is one form such that $dA=2J$ ($J$ is the Kahler form of
$CP^3$).
By comparing the above result with the conventional reduction formula,
we obtain type IIA background $AdS_4\times CP^3$
with 1-form and 3-form RR potentials as follows (setting
$\al=1$) \cite{ABJM,NPW}.
\begin{align}\label{CP3}
  ds^2 &= {\tilde R}^2(ds_{AdS_4}^2 + 4ds_{CP^3}^2),\no
  ds_{CP^3}^2
  &=d\zeta^2+\cos\zeta^2\sin^2\zeta\left(d\psi+\f{\cos\theta_1}{2}d\vp_1-
    \f{\cos\theta_2}{2}d\vp_2\right)^2 \no
  &\quad +\f{1}{4}\cos^2\zeta\left(d\theta_1^2+\sin^2\theta_1
    d\vp_1^2\right)+\f{1}{4}\sin^2\zeta(d\theta_2^2+\sin^2\theta_2d\vp_2^2),\\
  C^{(1)} &= \f{k}{2}\left[(\cos^2\zeta-\sin^2\zeta)d\psi +\cos^2\zeta
    \cos\theta_1 d\vp_1+ \sin^2\zeta \cos\theta_2 d\vp_2\right],\no
  C^{(3)} &= -\f{k{\tilde R}^2}{2}r^3\sin\t dt\we d\t\we d\vp,\no
  e^{2\phi}&= \f{4{\tilde R}^2}{k^2},\qquad  {\tilde R}^2=\f{R^3}{4k}=\pi\s\f{2N}{k}.
  \nonumber
\end{align}
Again, this background preserves 24 supersymmetries and its
holographic dual is recently argued to be the three dimensional
$\CN =6$ Chern-Simons theory (ABJM theory) \cite{ABJM}.

\section{Fuzzy Sphere and Dual Giant Gravitons}

\subsection{Dual Giant Gravitons in M-theory}
Consider a dual giant graviton expanding spherically in $AdS_4$ and
rotating in the $y$ direction of $S^7$ given in (\ref{S7}). We take
the world-volume coordinates of the M2-brane as
\begin{align}
  \sigma_0 \equiv \tau = t,\quad \sigma_1 = \t, \quad \sigma_2 = \vp,
\end{align}
and assume the transverse coordinates except $y$ does not depend on
the world-volume coordinates, and $y$ depends on $t$ as
\begin{align}
  y= y(t).
\end{align}
Remember that this coordinate $y$, defined in (\ref{S7}), is the
diagonal part of $\xi_i\ \ (i=1,2,3,4)$ in (\ref{corde}) and we fix
both the non-diagonal part of $\xi_i$ and the values of
$(\alpha,\beta,\gamma)$ to be constant. The action for this ansatz
is given by the DBI action and the Chern-Simons term
\begin{align}
  S &= -T_{2}\int d^3\sigma \s{-\det(P[G]_{ij})} - T_{2}\int P[C^{(3)}],\no
  &= - \f{\pi R^3T_{2}}{2}\int dt \left[ r^2\s{1+r^2-4{\dot y}^2} - r^3\right],
\end{align}
where $i,j$ and $P[\dots]$ denote the world-volume coordinates and
the pull-back on the world-volume respectively. The tension is given
by $T_2=\f{1}{(2\pi)^2}$ in our convention. The conserved momentum
conjugate to $y$ becomes
\begin{align}\label{Mom}
  P_{y} = 2\pi R^3T_{2} \f{r^2 \dot y}{\s{1+r^2 - 4{\dot y}^2}},
\end{align}
and the Hamiltonian becomes using $P_y$
\begin{align}
  H &= P_y\dot y - L\no
  &= \f{\pi R^3 T_2}{2}\left[ \s{1+r^2}\s{r^4 + \f{P_y^2}{(\pi R^3T_2)^2}}
  -r^3\right].
\end{align}
Thus we obtain the solution satisfying $\p H/\p r=0$ as
\begin{align}
  r=0, \qquad \s{\f{P_y^2}{(\pi R^3T_2)^2}}.
\end{align}
The former is a graviton and the latter is a giant graviton solution.
Substituting these into (\ref{Mom}), we can determine the dependence
on $t$ of $y$ as $y=\f{t}{2}$ or equally,
\begin{align}
  \xi_1=\xi_2=\xi_3=\xi_4 = \f{t}{2}.
\end{align}
Then the graviton and giant graviton rotate with the velocity of
light and have equal energies\footnote{The factor $1/2$ of right
hand side comes from the difference of the radius between $AdS_4$
and $S^7$; $2R_{AdS_4}=R_{S^7}=R$.}
\begin{align}
E=\f{P_y}{2}=\f{1}{2}(J_1+J_2+J_3+J_4), \label{boundf}
\end{align}
where $J_i=P_{\xi_i}$ are the momenta in the $\xi_i$ direction.

For example, if only one of $J_i$ is non-vanishing, this relation
(\ref{boundf}) shows the giant graviton is $\f{1}{2}$ BPS state
\cite{GMT} (for the properties of the supersymmetric states in the
dual $CFT_3$ see e.g.\cite{BhMi}). In the same way we can have $1/4$
and $1/8$ BPS dual giants depending on the values of
$(\ap,\beta,\gamma)$ as we will show in section 4. The result is
summarized in Table.\ref{tablenp}.

\subsection{Fuzzy Sphere in $AdS_4$}
Now we would like to show that dielectric D2-branes \cite{My} (or
fuzzy spheres) can be realized in the $AdS_4\times CP^3$ background
(\ref{CP3}). It expands spherically in $AdS_4$ and we take the
world-volume coordinates of D2-brane as
\begin{align}\label{D2WV}
  \sigma_0 \equiv \tau = t, \sigma_1 = \t, \sigma_2 = \vp,
\end{align}
and introduce the $U(1)$ field strength, describing the $M$ D0-brane
charge
\begin{align}
  F = \f{M}{2}\sin\t d\t\we d\vp.
\end{align}
Assuming the solution does not depend on $t$, the action for D2-brane
becomes
\begin{align}
  S = &= - T_2\int d^3\sigma e^{-\phi}\s{-\det (P[G]_{ij} + 2\pi
    F_{ij})} -T_2\int P[C^{(3)}]\no
  &= -2\pi T_2\,kR^2 \int dt \left[\s{r^2+1}\s{r^4+\f{\pi^2M^2}{R^4}}-r^3\right].
\end{align}
Since this Lagrangian consists of only potential term, the solution
is easily obtained by minimizing the Hamiltonian $H$
 leading to
\begin{align}
  r = \f{\pi M}{R^2},\qquad H = 2\pi^2T_2\, kM.
\end{align}
In our convention we have $T_2=\frac{1}{(2\pi)^2}$ and thus we find
the energy of the fuzzy sphere
\begin{align} E=\frac{kM}{2},\ \ \ \ \left(r=M\s{\f{k}{2N}}~\right).
\label{fuzzye}
\end{align}

The above fuzzy sphere configurations composed of the D2-D0 bound
states descend from the dual giant M2-branes in $AdS_4\times
S^7/Z_k$, which is rotating in the $\partial_{y}$ direction. This is
easily seen by taking the $Z_k$ orbifold $y\to y+\f{2\pi}{k}$ of the
dual giant in section 3.1. Indeed, the sum of the R-charges is
quantized as $J_1+J_2+J_3+J_4\in kZ$ due to the orbifold projection,
leading to the agreement between (\ref{fuzzye}) and (\ref{boundf}).

The fuzzy sphere D2-branes are parameterized by its position in
$CP^3$ in the IIA description. Some of them preserve a half
supersymmetries (12 SUSYs) when only one of $J_i\ \ (i=1,2,3,4)$ is
non-zero as summarized in the $k>2$ case of the Table.\ref{tablenp}.
Therefore, it is very interesting to consider its dual operator in
the ABJM theory (for a recent study of operators in ABJM theory dual
to giant gravitons refer to \cite{BeTr}) . When the number of
D0-branes is $M$, the energy or conformal dimension is given by
$E=\f{kM}{2}$ and the baryon charge is $kM$. Thus it should be made
of $k$ scalar fields of the matter fields
$(A_1,A_2,\bar{B}_1,\bar{B}_2)$. In order to have a gauge invariant
operator we need a contribution from the flux sector or the Wilson
line in the sense of \cite{ABJM}. It will be an interesting future
direction to pursuit the dual CFT operator in detail.

\section{Spinning Dual Giant Gravitons in M-theory}

As we have shown in the previous section the dual giant gravitons in
M-theory on $AdS_4\times S^7/Z_k$ are equivalent to the dielectric
D2-branes (fuzzy spheres) in type IIA string on $AdS_4\times CP^3$.
Partly motivated by the problem raised in the introduction
summarized in Fig.\ref{fuzzyspin}, we would like to move on to a
more non-trivial example by introducing a non-vanishing spin in the
$AdS_4$ direction. Another motivation for this is to understand the
$\f{1}{16}$ BPS states in $AdS_4\times S^7$. In this section we will
analyze the M-theory description of the spinning dual giant
gravitons and construct exact solutions by solving the BPS equation.
In the next section we reduce the solution to the fuzzy torus in
type IIA string.

\subsection{Spinning Giant Graviton Ansatz}
We assume the following ansatz\footnote{The unspecified parts are
the same as the non-spinning dual giant. If we write all components
explicitly in the coordinate system (\ref{corc}) , then we have
$\xi_1=\xi_2=\xi_3=\xi_4=w \vp+\omega t$ and $(\ap,\beta,\gamma)$
are fixed to be constant.} of the spinning dual giant graviton in
$AdS_4\times S^7/Z_k$
\begin{align}\label{RGG}
r=r(\theta),\ \ \ \ y=w \vp+\omega t,
\end{align}
where $w\in \f{Z}{k}$ is the winding number equivalent to
fundamental string charge (remember that $y$ is compactified as
$y\sim y+\f{2\pi}{k}$), while non-zero constant $\omega$ leads to
the D$0$-brane charge after the Kaluza-Klein reduction.

Then the M$2$-brane world-volume theory becomes
\begin{align}
S=-T_2\int dtd\theta d\vp {\cal{L}},
\end{align}
where
\begin{align}
\f{8}{R^3}{\cal{L}}=\s{\left(\f{r'^2}{1+r^2}+r^2\right)\left(
r^2(1+r^2)\sin^2\theta + 4w^2(1+r^2) - 4\om^2r^2\sin^2\t\right)}-
r^3\sin\theta.
\end{align}

\subsection{Supersymmetry Condition (generic case)}\label{sc:SC}
We would like to find M2-brane configurations that preserve some
supersymmetries. The projection operator of supersymmetries in the
presence of M2-branes is given by
\begin{align}\label{GM}
\Gamma=\f{1}{3!\s{-\det P[G]}}\ep^{ijk}\partial_i X^\mu \partial_j
X^\nu
\partial_k X^\rho \Gamma_{\mu\nu\rho},
\end{align}
which always satisfies $\Gamma^2=1$. We will analyze the
supersymmetry conditions closely following the general strategy in
\cite{MaSu}, where non-spinning dual giants in $AdS_5\times S^5$ has
been studied. In this subsection we assume that the M2-brane is
situated at a generic point of $(\ap,\beta,\gamma)$ in the
coordinate system (\ref{corc}). In section \ref{susyd}, we will
discuss the issue that the number of preserved supersymmetries is
enhanced at particular values of $(\ap,\beta,\gamma)$.

To examine the supersymmetry condition, let us notice $\Gamma_y=R
\sum_{i=1}^4\mu_i\gamma_{6+i}$. Instead we will write
$\Gamma_y=R\gamma_y$ for simplicity, but notice that $\gamma_y$
depends on the coordinate $\alpha,\beta$ and $\gamma$. In the end we
find (we define $r\equiv\sinh \rho$)
\begin{align}
  \Gamma & =\f{1}{3!\s{-\det P[G]}}(\Gamma_0+ \omega
  \Gamma_y)(\Gamma_2+r'\Gamma_1) (\Gamma_3+w\Gamma_y)  + (\mbox{perm})\no
  &=\f{R^3}{8\cdot 3!\s{-\det P[G]}}\left(\s{1+r^2}\ \gamma_0 + 2 \omega
    \gamma_y\right)\left(r\gamma_2+\f{r'}{\s{1+r^2}}\gamma_1\right)
  (r\sin\theta \gamma_3 +2w \gamma_y) + (\mbox{perm})\no
  &=\f{R^3}{8\cdot\s{-\det P[G]}}\Biggl[\s{1+r^2}r^2\sin\theta\gamma_{023}+
  2rw\s{1+r^2}~\gamma_{02y}+r'r\sin\theta~\gamma_{013}+2r'w\gamma_{01y}\no
  &\qquad +2\omega r^2\sin\theta~\gamma_{y23}+\f{2\omega
    r'r\sin\theta}{\s{1+r^2}}~\gamma_{y13} \Biggr]\no
  &=\f{R^3}{8\cdot\s{-\det P[g]}}(\rho'\g_1 + \sinh \rho \g_2)
  \Biggl[ 2(\om\sinh \rho\sin\t\g_{3} - w
  \cosh \rho \g_{0})\g_y - \sinh \rho\cosh \rho\sin\t \g_{03} \Biggr],
\end{align}
and
\begin{align}
  \s{-\det P[G]} = \f{R^3}{8}\s{(\rho'^2 + \sinh^2 \rho)(\sinh^2
  \rho \cosh^2 \rho\sin^2\t +
    4w^2\cosh^2 \rho - 4\om^2\sinh^2 \rho\sin^2\t)}.
\end{align}
The M2-brane preserves a fraction of supersymmetries specified by
\begin{align}\label{kappa}
(\Gamma+1)\ep=0,
\end{align}
out of the total 24 (or 32) Killing spinors when $k=1,2$ (or when
$k>2$). This requirement is highly non-trivial since both $\Gamma$
and $\ep$ depend on the coordinates. Multiplying $(\rho'\g_1 + \sinh
\rho \g_2)$ from the left, (\ref{kappa}) becomes
\begin{align}\label{kappa2}
  &\left[ \s{s}\big\{\sinh \rho\sin\t\g_3(2\om\g_y + \cosh \rho
    \g_0)- 2w\cosh \rho\g_0\g_y\big\} + \rho'\g_1 + \sinh \rho
    \g_2\right] \CM\ep_0 = 0,\no
  &\qquad \s{s} \equiv \s{\f{\rho'^2 + \sinh^2 \rho}{\sinh^2
  \rho\cosh^2 \rho\sin^2\t +
      4w^2\cosh^2 \rho -4\om^2 \sinh^2 \rho \sin^2\t}}.
\end{align}

To find the configuration preserving the supersymmetry, we suppose the
following conditions so that the above projection does not
depend on $\a,\b,\g$ when we move $\CM$ to the left
\begin{align}\label{1/16}
  &(\g_{47}-\g_{10}\hat\g)\ep_0=0, \qquad (\g_{58}-\g_{10}\hat\g)\ep_0=0,\no
  &(\g_{69}-\g_{10}\hat\g)\ep_0=0.
\end{align}
These three conditions are not independent because of the relation
$\g_{012345678910}\ep_0=\ep_0$ and the two of them become
independent. Under this conditions, we can move the matrix $\CM$ to
the left using the identities (\ref{GammaId}) given in the appendix
\begin{align}\label{Left}
   \CM\Bigg[ &e^{t\g_0\hat\g}\sinh \rho \Big\{\!\cosh \rho
   \s{s}\big\{\! 2w\cos\t\g_{2310}+
   \sin\t(\sin\vp \g_{2}\! -\! \cos\vp\g_{3})\g_{10}(2\om\! -\!\g_{010}\! -\! 2w\g_1)
   \big\}
   \!-\! \rho'\hat\g\Big\}\no
   &+\big\{ (\rho'\cosh \rho\cos\t - \sinh \rho \sin\t)\g_1
   -2w\cosh^2 \rho \s{s}\g_{010} -
   2\om \sinh^2 \rho \sin^2\t \s{s}\g_{0110}\big\} \no
   & + (\cos\vp\g_2 +\sin\vp\g_3)\big\{ \rho'\cosh \rho \sin\t
   + \sinh \rho \cos\t -2\om
   \sinh^2 \rho\sin\t\cos\t\s{s}\g_{010}\big\}
   \Bigg]  \ep_0 =0
\end{align}
To drop the time dependence in the first line, the following two
conditions are needed
\begin{align}\label{Susy}
  &(2\om -\g_{010} - 2w\g_1)\ep_0=0,\no
  &\g_{023}\left( -2w\cosh \rho \cos\t \s{s}\g_{010} + \rho'\g_1 \right)\ep_0=0.
\end{align}
Since $\g_1$ and $\g_{010}$ commute with each other, we can
simultaneously diagonalize these matrices. Denoting the eigenvalues
of these matrices by $\eta_1(=\pm 1)$ and $\eta_2(=\pm 1)$
respectively, the above equations are solved by
\begin{align}
  2\om &= \eta_2(2\eta_1\eta_2w + 1),\no
  \rho' &=
  -\f{2\eta_1\eta_2w\sinh \rho \cosh \rho \cos\t}{(
     \sinh^2 \rho-2\eta_1\eta_2 w)\sin\t}  \qquad (\eta_1\eta_2 w >0, \quad
  \sinh^2\rho < 2|w|) \label{GRin} \\
   &= \f{2\eta_1\eta_2w\sinh \rho \cosh \rho \cos\t}{(\sinh^2 \rho -
    2\eta_1\eta_2 w)\sin\t}  \qquad (\mbox{otherwise}) \label{GRing}
\end{align}
Moreover, we can check whether the second and third line of
(\ref{Left}) vanish or not, and we eventually find that only the
latter case \ie (\ref{GRing}) satisfies it. The conditions
(\ref{1/16}) and (\ref{Susy}) for the Killing spinor tell that this
solution is (at least) $\f{1}{16}$ BPS.

The solution (\ref{GRing}) is classified into the following two
types:
\begin{itemize}
\item For $\eta_1\eta_2 w < 0$, $\rho$ decreases monotonically
from $\rho(0)=\infty$ to $\rho\left(\f{\pi}{2}\right)$ with $\t$
varying from $0$ to $\f{\pi}{2}$.
\item For $\eta_1\eta_2 w > 0$ and $\sinh^2\rho \ge 2|w|$, $\rho$ increases from
$\rho(\t_0)$ to $\rho(\f{\pi}{2})$ with $\t$ varying from $\t_0$ to
$\f{\pi}{2}$, satisfying $\sinh^2\rho(\t_0) = 2\eta_1\eta_2 w$.
\end{itemize}
We can extend these solutions beyond $\theta=\f{\pi}{2}$ in a
symmetric way under the $Z_2$ action $\theta\to \pi-\theta$. In the
first case, we will have a sphere with two spikes attached (we will
analyze this in detail later, see Fig.\ref{GS3d}). In the second
case, the M2-brane world-volume naively ends at $\theta=\theta_0$.
This is due to the assumption that $r$ is single-valued function of
$\t$ in deriving (\ref{kappa}). Actually, the correct interpretation
of the second case turns out to be a toroidal world-volume (as we
will examine in detail later, see Fig.\ref{GT3d}), where $r$ becomes
double-valued. That is to say, the above description covers only a
half part of the whole configuration. If we choose the appropriate
world-volume coordinate $\chi$ corresponding to the one cycle of the
torus, instead of $\t$, the relation between the $\kappa$-projection
gamma matrices (\ref{GM}) becomes
\begin{align}
  \G(t,\chi,\vp) = \f{\p\t / \p\chi}{|\p\t / \p\chi|}\G(t, \t, \vp).
\end{align}
Therefore, in second case, the sign flips at $\t = \t_0$ and we must
use
\begin{align}
  (-\G +1)\ep = 0
\end{align}
instead of (\ref{kappa}) to extend the above configuration.
We can solve this equation similarly and obtain the supersymmetric solution
which connects the above solution as
\begin{align}
  \rho'  &= \f{2\eta_1\eta_2w\sinh \rho \cosh \rho \cos\t}{(\sinh^2 \rho
  -2\eta_1\eta_2 w)\sin\t} \qquad (\eta_1\eta_2 w > 0,\quad \sinh^2\rho \le
  2|w|).
\end{align}

In summary, we find that a spinning dual M2-giant in $AdS_4\times
S^7$ (or $AdS_4\times S^7/Z_2$) becomes
 at least $\f{1}{16}$ BPS state if the following
BPS equation is satisfied
\begin{align}\label{Sol}
  2\om &= \eta_2(2\eta_1\eta_2w + 1),\no
  \rho'  &= \f{2\eta_1\eta_2w\sinh \rho \cosh \rho \cos\t}{(\sinh^2 \rho -
    2\eta_1\eta_2 w)\sin\t}.
\end{align}
We can also rewrite this BPS equation in terms of the coordinate $r$
\begin{align}\label{Solr}
\f{dr}{d\theta}=2\eta_1\eta_2 w\f{r(1+r^2)}{r^2-2\eta_1\eta_2
w}\cdot\f{\cos\theta}{\sin\theta}.
\end{align}
As we will explain in the next subsection, it becomes $\f{1}{4}$ or
$\f{1}{8}$ BPS states if we choose specific values of $\ap,\beta$
and $\gamma$. We will leave the details of the analysis of
supersymmetry enhancement to the section \ref{susyd}.

A spinning dual M2-giant in $AdS_4\times S^7/Z_k\ \  (k>2)$ can be
treated similarly by just taking the orbifold $y\sim y+\f{2\pi}{k}$.
This orbifold projection kills some of the supersymmetries. In fact,
in the generic case, no supersymmetries will be left after the
orbifolding. However, as we will explain in section \ref{susyd}, for
specific values of $\ap,\beta$ and $\gamma$, there will be remaining
supersymmetries, leading to $\f{1}{4}$ or $\f{1}{12}$ BPS states in
$AdS_4\times S^7/Z_k$.

\subsection{BPS Equation from Bogomolnyi Bound}

In this section, we derive the supersymmetric equation (\ref{Sol})
or (\ref{Solr}) from another approach of the Bogomolnyi bound. We
can show the DBI part of the action is rewritten as follows
\begin{align}
  &\f{8}{R^3}\s{-P[G]}=\s{\left(\f{r'^2}{1+r^2}+r^2\right)\left(r^2(1+r^2)
      \sin^2\theta+4w^2(1+r^2)-4\omega^2r^2\sin^2\theta\right)} \no
  &=\s{\left(2w\f{d(r\cos\theta)}{d\theta}+\eta r^3\sin\theta\right)^2+
    \f{(r^2+2w)^2\sin^2\theta}{1+r^2}\left(\f{dr}{d\theta}-
      \f{2\eta wr(1+r^2)\cos\t}{(r^2-2\eta w)\sin\t}\right)^2},
\end{align}
imposing the relation\footnote{The over role $\pm$ sign is equal to
$\eta_1$.}\begin{align}
  \omega=\pm\left(w+\f{\eta}{2}\right),
\end{align}
where we can allow both signs $\eta=\pm 1$, corresponding to the
freedom $\eta_1\eta_2=\pm 1$ in (\ref{Solr}). If the BPS equation
(\ref{Solr}) or equally
\begin{align}\label{Solrr}
\f{dr}{d\theta}=2\eta w\f{r(1+r^2)}{r^2-2\eta
w}\cdot\f{\cos\theta}{\sin\theta},
\end{align}
is satisfied (setting $\eta=\eta_1\eta_2$),  then the total
Lagrangian
\begin{align}
\f{8}{R^3}{\cal{L}}=\f{8}{R^3}\s{-P[G]}\pm r^3\sin\theta =\mp 2\eta w
\f{d(r\cos\theta)}{d\theta}, \label{lagc}
\end{align}
successfully becomes a total derivative as required from the
ordinary Bogomolnyi bound argument. At the same time, this
guarantees that the solution to the BPS equation (\ref{Solrr})
satisfied the equation of motion. Notice that the coefficient in
front of $r^3\sin\theta$ represents whether we consider a M2-brane
or an anti M2-brane. By plugging the explicit solutions, we can
check that in the Lagrangian (\ref{lagc}), the $r^3\sin\theta$ term
from the square root always cancels out the $r^3\sin\theta$ term
from the coupling to 3-form field. In this way, we rederived the BPS
equation (\ref{Sol}) or (\ref{Solr}) from the Bogomolnyi bound
argument.

We would also like to compute energy and angular momenta. There are
five different angular momenta: $J_i=P_{\xi_i}\ \ (i=1,2,3,4)$ and
$S=P_\vp$. In the $AdS_4/CFT_3$ viewpoint $J_i$ is the R-charge and
$S$ is the spin inside the $AdS_4$. They are obtained by taking the
functional derivative as $P_{\vp}= -T_2\int d\theta d\vp \f{\partial
{\cal{L}}} {\partial \dot{\vp}}$ and we find the linear relation
\begin{align}
  J_i=\f{\mu_i^2}{w}S,
\end{align}
where $\mu_i=(\sin\ap,\cos\ap\sin\beta,\cos\ap\cos\beta\sin\gamma
,\cos\ap\cos\beta\cos\gamma)$. If we define
$J=\sum_{i=1}^4J_i~(=P_y)$, then we find the relation $wJ=S$.

The energy is given by
\begin{align}
E=\int d\theta d\vp \left[\sum_{i=1}^4
P_{\xi_i}\dot{\xi_i}-(-T_2){\cal L} \right] =
\f{\eta_1}{2}(2S+J_1+J_2+J_3+J_4) + T_2\int d\theta d\vp {\cal L},
\label{boundl}
\end{align}
where $\eta=\pm 1$ is the sign introduced in (\ref{GRing}).
 Notice that this relation almost saturates the BPS bound expected
 from the dual $CFT_3$ since the final term
 $\int d\theta d\vp {\cal L}$ is a total
derivative (\ref{lagc}).

\subsection{Analytical Solutions to BPS Equation}

Now let us solve the BPS equation (\ref{Solrr}) and determine the
shape of M2-brane. We can analytically solve the differential
equation and finally we get
\begin{align}
  \sin\theta=A\cdot r^{-1}(1+r^2)^{\f{1}{2}+\f{1}{4\eta w}},
  \label{bpssol}
\end{align}
where $A$ is a constant. If we instead employ the coordinate $\rho$
we obtain
\begin{align}
  \sin\theta=A\cdot\frac{(\cosh\rho)^{1+\frac{1}{2\eta w}}}{\sinh \rho}.
\end{align}\\

Since the solution is $Z_2$ symmetric $r(\theta)=r(\pi-\theta)$, we
have only to discuss the behavior of the function $r(\theta)$ for
$0\leq \theta\leq \f{\pi}{2}$. It is easy to see that when $\eta
w<0$ there are only one value of $r$ which satisfies (\ref{bpssol}),
while when $\eta w>0$ there are two such solutions. Accordingly, we
have two different types of the M2-brane shape depending on the sign
of $\eta w$.
\\

\subsubsection{Case 1: $\eta w< 0$ (Giant Spike)}

\begin{figure}[htbp]
  \begin{center}
    \includegraphics[height=7cm]{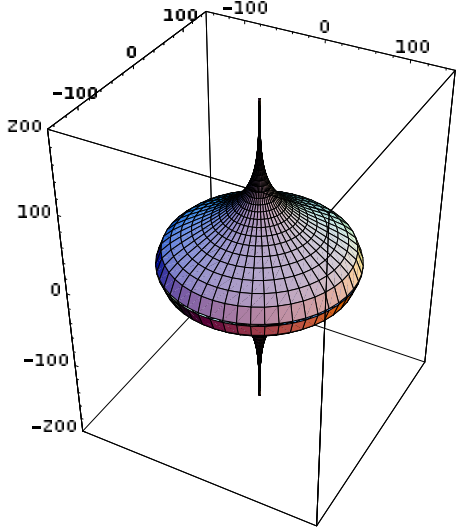}
  \end{center}
  \caption{The giant-spike solution described in the three
  dimensional space $(r,\theta,\vp)$ (we assumed $|w|=10$
    and $A=3$). This corresponds to the sign $\eta w<0$.
    We multiplied the factor 10 with the horizontal
    coordinates.}
  \label{GS3d}
\end{figure}

In this case, $r(\theta)$ is monotonically decreasing function when
$0\leq \theta\leq \f{\pi}{2}$. It satisfies $r(0)=\infty$ and
$r(\f{\pi}{2})=r_0$, where the constant $r_0>0$ is related to the
integration constant $A$ via
$A=r_0(1+r_0^2)^{-\f{1}{2}+\f{1}{4|w|}}$. The shape of this M2-brane
world-volume is plotted in Fig.\ref{GS3d}. It looks like a sphere
with two spikes attached. The spikes are actually a cylinder with
radius $\f{R}{k}$ winding $wk$ times in the $y$ direction. Thus it
can be regarded as a bound state of a (non-spinning) dual M2 giant
and $wk$ tubular M2 branes\footnote{This tubular part locally looks
similar to the M-theory lift of the supertube discussed in
\cite{HyOh}.}.

Its energy is calculated from (\ref{boundl}) as follows (we choose
$\eta_1=+1$)
\begin{align}
E-S-\f{1}{2}(J_1+J_2+J_3+J_4)&=\f{R^3}{8}T_2 \int d\theta d\vp~
2w[r\cos\theta]'\no &= \pi w R^3 T_2 (r_{\infty}-r_0),
\end{align}
where $r_\infty$ presents the infinitely large value of $r$ at
$\theta=0$ and $\theta=\pi$. Indeed the infinitely large
contribution of right-hand side precisely coincides with the
infinitely large mass of $kw$ M2-branes which wrap on the circle
$\partial_{y}$ and which extend in the $r$ direction. We can
explicitly express $J=\sum_{i=1}^4 J_i$ as follows
\begin{align}
J&=\f{R^3}{2}\omega T_2\int d\theta d\vp r^2\sin^2\theta
\s{\f{\f{r'^2}{1+r^2}+r^2}{r^2\sin^2\theta(1+r^2)-4\omega^2
r^2\sin^2\theta+4w^2(1+r^2)}}\no &=2\pi R^3\omega
T_2\int_{r_0}^{\infty} dr \f{r^2\sin^2\theta}{2w(1+r^2)\cos\theta}.
\end{align}

\subsubsection{Case 2: $\eta w> 0$ (Giant Torus)}

In this case, the values of $r$ in (\ref{bpssol}) is restricted to a
certain range $r_-\leq r\leq r_+$ because the function
$B(r)=r^{-1}(1+r^2)^{\f{1}{2}+\f{1}{4|w|}}$ takes its minimum at
$r=\s{2w}$ and satisfies $B(0)=B(\infty)=\infty$. The values of
$r_{\pm}$ are given by the two solutions to
$Ar_0^{-1}(1+r_0^2)^{\f{1}{2}+\f{1}{4|w|}}=1$. It is clear that we
need to require $A<1/B(\s{2w})$ in order to have any solution.

\begin{figure}[htbp]
  \begin{center}
    \includegraphics[height=5cm]{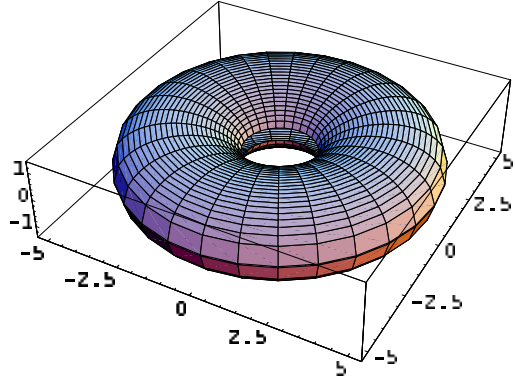}
  \end{center}
  \caption{The giant-torus solution corresponding to the sign $\eta
  w>0$ (we assumed $|w|=10$
    and $A=0.8$).}
 \label{GT3d}
\end{figure}

The values of $\theta$ is restricted to the range
$\theta_0\leq\theta\leq\pi-\theta_0$ such that $r(\theta_0)=\s{2w}$.
For a given $\theta$ there are two values of $r$ which satisfy
(\ref{bpssol}). Thus we can conclude that the world-volume of this
M2-brane is topologically a torus. An explicit shape of this
M2-brane world-volume is plotted in Fig.\ref{GT3d}.

In this case, the total derivative term in (\ref{boundl}) just
vanishes since the world-volume has no boundary. Thus we obtain the
standard BPS formula (choosing $\eta_1=+1$)
\begin{align}
E=S+\f{1}{2}(J_1+J_2+J_3+J_4)=\omega J,
\end{align}
and
\begin{align}
J=2\pi R^3\omega T_2\int_{r_-}^{r_+} dr
\f{r^2\sin^2\theta}{2w(1+r^2)\cos\theta}.
\end{align}
 It is interesting to note that
 if we fix $J$ and take $w$ to be large, then we can realize a ring-like
object.

\subsection{Enhanced Supersymmetries} \label{susyd}
In the previous subsection, we showed that the solution which
satisfies (\ref{Solr}) preserves at least two (\ie $\f{1}{16}$ BPS)
of the total thirty-two supersymmetries in $AdS_4\times S^7$. Though
for general values of $\a, \b$ and $\g$, the spinning dual giant is
a $\f{1}{16}$ BPS state, we can see that it enhances to $\f{1}{4}$
or $\f{1}{8}$ BPS states for specific values of $\a, \b$ and $\g$.
In this subsection we will examine these details of supersymmetries
in both $AdS_4\times S^7$ and its orbifold $AdS_4\times S^7/Z_k$.
The result is summarized in the Table.\ref{tablenp} (non-spinning
case) and Table.\ref{tablesp} (spinning case).

It is convenient to define the $\pm 1$ eigenvalues
$\eta_1,\eta_2,s_1,s_2,s_3$ and $s_4$ of the commuting matrices
$\gamma_1,\gamma_{01\!0},\gamma_{47},\gamma_{58},\gamma_{69}$ and
$\hat{\gamma}\gamma_{10}$ for a given spinor $\ep_0$ as we have
already did so in previous sections
\begin{align}
  & (\gamma_1-\eta_1)\ep_0=0, \qquad (\gamma_{01\!0}-\eta_2)\ep_0=0,\no
  &(\g_{47}-is_1)\ep_0=0,  \qquad (\gamma_{58}-is_2)\ep_0=0,\no
  &(\g_{69}-is_3)\ep_0=0,  \qquad
  (\hat{\gamma}\gamma_{10}-is_4)\ep_0=0.
\end{align}
Since the 11D spinor $\ep_0$ is chiral, we impose
$\gamma_{01234567891\!0}\ep_0=\ep_0$. This leads to the relation
$s_1s_2s_3s_4=1$.

\subsubsection{Enhanced Supersymmetries in $AdS_4\times S^7\ (\mbox{or}\
S^7/Z_2)$}
For example, if we set $\a=\b=\g=0$ (called case (a)), we do not
need to require any of the constraints in (\ref{1/16}) because
$\gamma_y=\gamma_{10}$. Thus in this case, it becomes $\f{1}{4}$ BPS
because we only need to fix the signs of $\eta_1$ and $\eta_2$. In a
similar way, when two out of $\a,\b$ and $\g$ are vanishing (called
case (b)), we get a $\f{1}{8}$ BPS state by requiring a further
constraint $s_1=s_4$, $s_2=s_4$ or $s_3=s_4$. For other generic
values (case (c)), it becomes $\f{1}{16}$ BPS as we already
mentioned.

Also notice that non-spinning case $w=0$ is special. In this case we
do not need to specify the value of $\eta_1$ as is clear from
(\ref{Susy}). Thus the number of preserved supersymmetries becomes
doubled as summarized in Table.\ref{tablenp}.

\subsubsection{Enhanced Supersymmetries in $AdS_4\times S^7/Z_k\ \ (k>2)$}
The $Z_k$ orbifold action on the spinor $\ep_0$ produces the phase
factor
\begin{align}
\ep_0\ \to\ e^{\f{\pi i}{k}(s_1+s_2+s_3+s_4)}\ep_0.
\end{align}
Thus for the M-theory on $AdS_4\times S^7/Z_k$ with $k>2$ (or IIA
string on $AdS_4\times CP^3$), there are 24 Killing spinors
corresponding to the choice $(+,+,-,-)$ and its permutations, which
satisfies the orbifold projection $\sum_{i=1}^4s_i\equiv 0\
(\text{mod}~2)$ as already mentioned.

The number of supersymmetries for spinning or non-spinning dual
giants can be analyzed in the same way as in the previous $k=1,2$
case. It is again summarized in Table.\ref{tablenp} and
Table.\ref{tablesp}.

\vspace{1cm}

\begin{table}[htbp]
  \begin{center}
    \begin{tabular}{|c|c|c|c|c|}
      \hline
      $k$ & Total SUSY & Case (a) & Case (b) & Case (c) \\ \hline
      $k=1,2$ & 32 & 16 & 8 & 4 \\ \hline
      $k>2$ & 24 & 12 & 4 & 0 \\
      \hline
    \end{tabular}
  \end{center}
  \caption{The number of supersymmetries of dual {\it non-spinning}
   giant gravitons
   in $AdS_4\times S^7/Z_k$. Its world-volume looks like a sphere as usual.
   The result of dielectric D2-branes (or fuzzy sphere) in
    type IIA string on $AdS_4\times CP^3$
    corresponds to the $k>2$ case. The case (a) is defined by
    $\ap=\beta=\gamma$. The case (b) is by $\ap=\beta$,
    $\beta=\gamma$ or $\ap=\gamma$. The case (c) is all the other possibilities.}
    \label{tablenp}
\end{table}

\begin{table}[htbp]
  \begin{center}
    \begin{tabular}{|c|c|c|c|c|}
      \hline
      $k$ & Total SUSY & Case (a) & Case (b) & Case (c) \\ \hline
      $k=1,2$ & 32 & 8 & 4 & 2 \\ \hline
      $k>2$ & 24 & 6 & 2 & 0 \\
      \hline
    \end{tabular}
  \end{center}
  \caption{The number of supersymmetries of dual {\it spinning}
  giant gravitons
    in $AdS_4\times S^7/Z_k$. Its world-volume looks like either a
    torus or sphere with two spikes. The result of spinning
    dielectric D2-branes
    (or fuzzy torus) in type IIA string on $AdS_4\times CP^3$
    corresponds to the $k>2$ case. The cases (a), (b) and (c) are
    defined in the same way as in Table.\ref{tablenp}.}\label{tablesp}
\end{table}

\section{Fuzzy Rings in $AdS_4$}

As we learned in section 3, we can reduce a spherical dual giant
graviton of M2-brane to a fuzzy sphere of a dielectric D2-brane in
$AdS_4\times CP^3$ by considering the orbifold $AdS_4\times
S^7/Z_k$. Thus it is intriguing to apply the same procedure to the
spinning dual giant.

By construction, it is rotating both in the $y$ and $\vp$ direction.
The angular momentum $J=P_y\equiv kM$ leads to $M$ units of the
D0-brane charge after the reduction to IIA string, while the winding
number $w$ corresponds to the $wk$ units of the F-string
charge\footnote{Remember that in the orbifold theory of $AdS_4\times
S^7/Z_k$ the winding number $w$ is fractionally quantized i.e.
$wk\in Z$.}. Therefore, the spinning dual giant constructed in the
previous section will be reduced to a bound state of a D2-brane, $M$
D0-branes and $wk$ F-strings. Since the F-strings and D0-branes
generate the electric and magnetic flux, the system has a
non-vanishing Poynting vector which produces non-vanishing angular
momentum (or spin) in the $\vp$ direction. Indeed its value is given
by $S=P_{\vp}=wkM$. Notice that the quantization of the angular
momentum is a consequence of the charge quantization of the F-string
and D0-brane. In this way, in order to obtain a non-vanishing spin,
the F-string charge is necessary.

One may notice that there are two different ways of attaching the
F-string to the fuzzy sphere: (i) F-strings which are attached at
north and south poles on the sphere and stretch toward the infinity
(ii) F-strings which connects between the two poles. Indeed, the BPS
equation in section 4 precisely leads to corresponding two
solutions: giant spike and giant torus.

The profile of this dielectric D2-brane is again given by
(\ref{bpssol}). The values of gauge fluxes can be computed by
rewriting\footnote{ Since we take Kaluza-Klein reduction fixing the
radius of $y$ direction, the M2-brane tension $T_2$ is equal to the
D2-brane tension $T_2$ shown in (\ref{M2D2}).} the action of a
D2-brane
  into that of a M2-brane \cite{To,Sc}
\begin{align}\label{M2D2}
  S_{DBI} &= -T_{2}\int_{D2}d^3x\ e^{-\phi}
  \s{-\det(P[G^{(IIA)}]_{ij} + 2\pi F_{ij})} \no
  &= -T_{2}\int_{D2}d^3x \left[ e^{-\phi}
  \s{-\det(P[G^{(IIA)}]_{ij} + e^{2\phi}a_ia_j)}
    + \pi \ep^{ijk}a_i F_{jk}\right]\\
  &= - T_{2}\int_{M2} d^3x \s{-\det(P[G^{(M)}]_{ij}
 + e^{\f{4}{3}\phi}\p_i \ti{y} \p_j \ti{y})},\nonumber
\end{align}
where we defined $\p_i \ti{y}=a_i$ in the final expression. The
integration over the D2-brane gauge field $A_i$ requires the
auxiliary vector field $a_i$ to be a total derivative. Notice also
that the metric in M-theory and the IIA string frame metric is
related via
$ds^2_{M}=e^{-\f{2}{3}\phi}ds^2_{IIA}+e^{\f{4}{3}\phi}(d\ti{y}+C)^2$.
We can check the equivalence of the first line and the second line
by integrating out $a_i$ explicitly\footnote{Here we use the
identity written in \cite{Sc}
\begin{align}
  -\det (P[G]_{ij} + 2\pi F_{ij}) &= (-\det P[G])(1+2\pi^2 F^2),\no
  -\det (P[G]_{ij} + e^{2\phi}t_it_j) &= (-\det P[G])
  (1+e^{2\phi}t_it^i).
\end{align}
}.

The relation between $F_{ij}$ and $\p_i \ti{y}$ is given by
\begin{align}
  e^{\f{4}{3}\phi}
  \s{-\det P[G^{(M)}]} \f{\p^i \ti{y}}{\s{1+e^{\f{4}{3}\phi}
  \p_i \ti{y} \p^i \ti{y}}} = -\pi\ep^{ijk}F_{jk}, \label{relaa}
\end{align}
where the left-hand side should be computed by using the M-theory
metric.

By plugging $\ti{y}=ky=k(\omega t+w\phi)$ (\ref{RGG}) to
(\ref{relaa}), we eventually obtain the electric and magnetic field
on the spinning dielectric D2-brane as follows\footnote{We used the
relation $\f{R^3}{4k}=\pi\s{\f{2N}{k}}$ in (\ref{CP3}).}
\begin{align}
& F_{t\theta}=-w\s{\f{2N}{k}}\cdot\f{r(1+r^2)}{|r^2-2\eta
w|\sin\theta},\no & F_{\theta\vp}=\omega
\s{\f{2N}{k}}\cdot\f{r^3\sin\theta}{|r^2-2\eta w|}.
\end{align}
Note that if we set $w=0$ and $\omega=\f{1}{2}$, then we correctly
reproduce $F_{t\theta}=0$ and $F_{\theta\vp}=\f{M}{2}\sin\theta$ as
discussed in section 3.2.

In the same way as the spinning dual giant, we have two different
types of the world-volume \ie a sphere with two spikes (see
Fig.\ref{GS3d}) and a torus
 (see Fig.\ref{GT3d}). The former case with
infinitely long spikes is naturally understood as a bound state of
the D2 fuzzy sphere and infinitely long fundamental strings. Since
it has an infinite energy, it is not appropriate for the candidate
of `rotating fuzzy sphere' raised in the introduction of this paper.
Instead, the fuzzy torus configuration will be a correct candidate
as explained in Fig.\ref{fuzzyspin}.

The supersymmetry of these D2-D0-F1 bound states is the same as the
analysis of spinning dual giants in $AdS_4\times S^7/Z_k$ with $k$
sufficiently large and is shown in Table.\ref{tablesp}.

Finally we would like to note that it is possible to make one of two
cycles of the torus very small so that it looks like a ring assuming
$w$ is very large. By considering its back-reacted geometry, this
fuzzy ring solution might suggest an existence of supersymmetric or
non-supersymmetric black ring solutions in a certain $AdS_4$
supergravity. However, the topological censorship\footnote{We are
very grateful to Mukund Rangamani for bring us this reference and
related discussions.} in $AdS_4$ \cite{GSWW} tells us that the
horizon topology should always be $S^2$ and not $T^2$. Therefore, it
is probable that it becomes a small black ring instead of a
macroscopic one, by taking into account higher derivative
corrections (see also \cite{IiSi} for a discussion of (small) black
rings in four dimension). It is also a very interesting future
problem to repeat a similar analysis for $AdS_5\times S^5$ and see
if we can realize a fuzzy ring. If such an object exists, it might
suggest supersymmetric black ring-like
objects\footnote{Supersymmetric black rings in standard gauge
supergravities such as the minimal gauged supergravity have been
investigated in \cite{KLR} and it has been shown that they do not
exist. However, in our case, the presence of 3-form gauge potential
in $AdS_4$ (or 4-form potential in $AdS_5$) is crucial for the
existence of the fuzzy ring. It might be possible for these extra
fluxes change the situation. To see if this is true or not will be
an interesting future problem.} in $AdS_5$. An evidence for
non-supersymmetric black ring solutions in $AdS$ has been given
recently in \cite{EmA}.

\section{Dual Giant Gravitons in Type IIA String on $AdS_4\times CP^3$}

Before we conclude this paper, we would like to mention spherical
D2-branes wrapped on $S^2$ in $AdS_4$ and orbiting the $CP^3$ as
they are other interesting BPS states dual to supersymmetric
operators in the ABJM theory. They can be regarded as dual giant
gravitons in type IIA string on $AdS_4\times CP^3$. They are also
obtained from the reduction of spherical dual giants in $AdS_4\times
S^7/Z_k$.

We take the world-volume coordinates of D2-brane as
\begin{align}
  \sigma_0 \equiv \tau = t, \sigma_1 = \t, \sigma_2 = \vp,
\end{align}
and consider a trial solution of the form
\begin{align}
  \psi = \psi (\tau),\quad \phi_1 = \phi_1 (\tau),\quad \phi_2 =
  \phi_2 (\tau),\quad
  \zeta, \t_1,\t_2 ,r = const.
\end{align}
Then, the D2-brane action is written as
\begin{align}\label{GGLag}
  S = -2\pi T_2\,kR^2 \int dt [r^2\D^{1/2}-r^3],
\end{align}
where $\D\equiv P[G]_{tt}$ is given by  (in the coordinate
(\ref{CP3}))
\begin{align}\label{Delta}
  \D &= 1+r^2 - 4\cos^2\zeta \sin^2\zeta\left( \dot\psi +
    \f{\cos\t_1}{2}\dot\vp_1 - \f{\cos\t_2}{2}\dot\vp_2 \right)^2 \no
  &\quad - \cos^2\zeta\sin^2\t_1\dot\vp_1^2 - \sin^2\zeta\sin^2\t_2\dot\vp_2^2.
\end{align}
Solving the above Lagrangian, we can obtain the spherical D2-brane
solution. For simplicity, we take $\zeta=\f{\pi}{4},\t_1=\t_2=0$ and
$\dot\vp_1=\dot\vp_2=0$. In this case, the Lagrangian (\ref{GGLag})
becomes
\begin{align}
  L = - a  \left[r^2\s{1+r^2-\dot\psi^2} - r^3\right],
\end{align}
where we denote $a=2\pi T_2\,kR^2$.
The momentum conjugate to $\psi$ is
\begin{align}
  P_\psi = \f{ar^2\dot\psi}{\s{1+r^2-\psi^2}}.
\end{align}
Using this, the Hamiltonian becomes
\begin{align}\label{Ham}
  H = P_\psi\dot\psi - L = a\left[\s{r^2+1}\s{r^4+\f{P_\psi^2}{a^2}} -
    r^3\right].
\end{align}
Thus the solution satisfying $\p H/\p r =0$ is
\begin{align}
  r=0,\qquad r^2=\f{P_\psi^2}{a^2}.
\end{align}
The former is graviton and the latter is giant graviton solution.
These have the equal energy $E = P_\psi=\f{1}{2}(J_1+J_2-J_3-J_4)$.
Since they do not rotate in the $y$ direction, the dual operators
will not have any baryon charge. Therefore they should be dual to be
singlet operators of the bi-fundamental matter fields such as the
symmetric polynomials of $(A_iB_j)$ as in the case of $AdS_5\times
S^5$ \cite{Ram}.

\section{Conclusion}

In this paper we presented an analytical description of spinning
dual giant gravitons in $AdS_4\times S^7$ and its orbifold. We
showed that its world-volume in $AdS_4$ looks like either a torus or
a sphere with two infinitely long spikes attached. We worked out the
number of supersymmetries which are preserved by this configuration.

We further reduced these M-theory BPS states to those in the type
IIA string by taking an orbifold. They are interpreted as spinning
dielectric D2-branes. Even though the world-volume of a non-rotating
dielectric D2-branes is given by a sphere, its topology should be
changed into a torus when we rotate it. This fuzzy torus is a bound
state of a D2-brane, D0-branes and F-strings and is spinning due to
the Poynting vector due to the presence of both electric and
magnetic gauge flux. If we imagine a dynamical process of increasing
angular momentum of a dielectric D2-brane, it is impossible for the
topology change to occur instantly. Therefore, it will be intriguing
to study the time-dependent process of developing an empty tube
inside the fuzzy sphere.

At the same time, our results offer new BPS objects in the $AdS_4$
backgrounds dual to the ABJM theory. It will be another interesting
future direction to explore their dual supersymmetric operators in
detail.

We also find that for an appropriate choice of parameters, the fuzzy
torus can degenerate into a fuzzy ring, which suggests the existence
of (possibly small) supersymmetric black rings in the $AdS_4$
spacetime. It may also be interesting to see if we can construct
similar dual giant gravitons in $AdS_5\times S^5$.

\vspace{0.5cm}

{\bf Notes Added:}
After publishing this paper, we realized that our 1/4-BPS solution 
is included in the general 1/4-BPS solutions constructed by
O. Lunin \cite{Lu}.
His method would be useful to count the microstates of the supersymmetric 
black hole.
We are grateful to O. Lunin for pointing out this fact.

\vspace{1cm} \centerline{\bf Acknowledgments}

\vspace{0.5cm}

We are very grateful to M. Hanada, T. Harmark, K. Hashimoto, H.
Kunduri, G. Mandal, S. Minwalla, N. Ohta, S. Ramgoolam, M.
Rangamani, and S. Terashima for valuable discussions and comments.
We would like to thank very much the Monsoon Workshop on string
theory at TIFR and the Summer Institute 2008 at Fujiyoshida, during
which important steps of this work have been taken. The work of TN
is supported by JSPS Grant-in-Aid for Scientific Research
No.19$\cdot$3589. The work of TT is supported by JSPS Grant-in-Aid
for Scientific Research No.18840027 and by JSPS Grant-in-Aid for
Creative Scientific Research No. 19GS0219.


\newpage
\appendix
\section{Killing spinors for $AdS_4 \times S^7$}\label{Ap:Killing}
Here we will construct the Killing spinors preserved in the $AdS_4
\times S^7$ background (\ref{metric}). The vielbeins are given by
\begin{align}
  \label{eq:vielbeins}
  &e^0=\f{R}{2}\s{1 + r^2}dt,\qquad
  e^1=\frac{Rdr}{2\s{1+\frac{r^2}{l^2}}},\qquad e^2=\f{Rrd\t}{2},\no
  &e^3=\f{Rr\sin\t d\ph}{2},\qquad e^4=Rd\a,\quad e^5=R\cos\a d\b,\no
  &e^6=R\cos\a \cos\b d\g, \qquad e^7=R\sin\a d\xi_1,\qquad
  e^8=R\cos\a\sin\b d\xi_2,\no
  &e^9=R\cos\a\cos\b\sin\g d\xi_3,\qquad
  e^{10}=R\cos\a\cos\b\cos\g d\xi_4,
\end{align}
and the spin connections defined as $de^\m + \om^\m_\n \we e^\n = 0$ becomes
\begin{align}
  \label{eq:spin_con}
  &\om^0_1=rdt,\qquad\om^1_2=-\s{1+r^2}d\t,\qquad
  \om^1_3=-\s{1+r^2}\sin\t d\ph,\no
  &\om^2_3=-\cos\t d\ph,\qquad  \om^4_5=\sin\a d\b,\qquad
  \om^4_6=\sin\a\cos\b d\g,\qquad \om^5_6=\sin\b
  d\g,\no
  &\om^4_7=-\cos\a d\xi_1,\qquad\om^4_8=\sin\a\sin\b
  d\xi_2,\qquad \om^5_8=-\cos\b d\xi_2,\qquad\no
  &\om^4_9=\sin\a\cos\b\sin\g d\xi_3,\qquad
  \om^5_9=\sin\b\sin\g d\xi_3,\qquad \om^6_9=-\cos\g
  d\xi_3,\no
  &\om^4_{10}=\sin\a\cos\b\cos\g d\xi_4,\qquad
  \om^5_{10}=\sin\b\cos\g d\xi_4,\qquad \om^6_{10}=\sin\g d\xi_4.
\end{align}
The supersymmetry transformation of the gravitino reads
\begin{align}
\label{eq:susy_tr}
&  \delta \Psi_\m \equiv D_\m\ep -\frac{1}{288}(\G_\m^{\n\lambda\rho\sigma} -
  8\delta_\m^\n\G^{\lambda\rho\sigma})F_{\n\lambda\rho\sigma}\ep, \\
& D_\m\psi \equiv(\p_\m
+\frac{1}{4}\om^{\n\rho}_\mu\g_{\n\rho})\psi, \nonumber
\end{align}
where $\ep$ and $\psi$ are Majorana fermions. Bosonic configuration
($\Psi_\m=0$) preserves the supersymmetry when $\delta\Psi_\m=0$ is
satisfied. Putting (\ref{eq:spin_con}) into (\ref{eq:susy_tr}), the
Killing spinor equation can be calculated as follows:

\noindent
$\underline{AdS_4}$
\begin{align}
  \label{eq:killing2}
&(\p_t+\frac{r}{2}\g_{01}+\frac{1}{2}\s{1+r^2}\g_0\hat\g)\ep=0,\no
& (\p_r + \frac{1}{2}\frac{1}{\s{1+r^2}}\g_1\hat\g)\ep=0,\no
&(\p_\t-\frac{1}{2}\s{1+r^2}\g_{12}+\frac{r}{2}\g_2\hat\g)\ep=0,\no
& (\p_\ph - \frac{1}{2}\s{1+r^2}\sin\t\g_{13}
-\frac{1}{2}\cos\t\g_{23}+\frac{r\sin\t}{2}\g_3\hat\g)\ep=0,
\end{align}\\
$\underline{S^7}$
\begin{align}
&(\p_\a-\frac{1}{2}\hat\g\g_4)\ep=0,\no
&(\p_\b+\frac{1}{2}\sin\a\g_{45}-\frac{1}{2}\cos\a\hat\g\g_5)\ep=0,\no
&(\p_\g
  +\frac{1}{2}(\sin\a\cos\b\g_{46}+\sin\b\g_{56}
)-\frac{1}{2}\cos\a\cos\b\hat\g\g_6)\ep=0,\no
&(\p_{\xi_1}-\frac{1}{2}\cos\a\g_{47}
-\frac{1}{2}\sin\a\hat\g\g_7)\ep=0,\no
&(\p_{\xi_2}+\frac{1}{2}(\sin\a\sin\b\g_{48}
-\cos\b\g_{58})
-\frac{1}{2}\cos\a\sin\b\hat\g\g_8)\ep=0,\no
&(\p_{\xi_3}+\frac{1}{2}(\sin\a\cos\b\sin\g\g_{49}
+\sin\b\sin\g\g_{59}-\cos\g\g_{69})
-\frac{1}{2}\cos\a\cos\b\sin\g\hat\g\g_9)\ep=0,\no
&(\p_{\xi_4}+\frac{1}{2}(\sin\a\cos\b\cos\g\g_{410}
+\sin\b\cos\g\g_{510}+\sin\g\g_{610})
-\frac{1}{2}\cos\a\cos\b\cos\g\hat\g\g_{10})\ep=0,
\end{align}
Solving these equations, we obtain the Killing spinor preserved by $AdS_4\times
S^7$
\begin{align}
\ep=e^{\f{\alpha}{2}\hat{\gamma}
\gamma_4}e^{\f{\beta}{2}\hat{\gamma}  \gamma_5}
e^{\f{\gamma}{2}\hat{\gamma} \gamma_6}e^{\f{\xi_1}{2}\gamma_{47}}
e^{\f{\xi_2}{2}\gamma_{58}}e^{\f{\xi_3}{2}\gamma_{69}}
e^{\f{\xi_4}{2}\hat{\gamma} \gamma_{10}}e^{-\f{\rho}{2}\gamma_1
\hat{\gamma}} e^{-\f{t}{2}\gamma_0 \hat{\gamma}}
e^{\f{\theta}{2}\gamma_{12}} e^{\f{\vp}{2}\gamma_{23}}\ep_0,
\end{align}
where $\ep_0$ is an arbitrary constant spinor, then we have 32
independent Killing spinors. That is to say, there are 32
supersymmetry for $AdS_4 \times S^7$ background.

\section{Useful Identities}
Below we present the useful gamma matrix identities used in section
\ref{sc:SC}
\begin{align}\label{GammaId}.
  \g_{0}\CM&=\CM e^{t\g_{0}\hat\g}\g_{0}\no
  \g_1\CM&=\CM (\cosh x \cos \t \g_1 + \cosh x\sin\t\cos\vp\g_2+\cosh
  x\sin\t\sin\vp\g_3 - \sinh x e^{t\g_0\hat\g}\hat\g)\no
  \g_{2}\CM&=\CM(-\sin\t\g_{1}+\cos\t\cos\vp\g_{2}+\cos\t\sin\vp\g_{3})\no
  \g_{3}\CM&=\CM(-\sin\vp\g_{2}+\cos\vp\g_{3})\no
  \g_{7}\CM&=\CM(\cosh xe^{t\g_{0}\hat\g} + \sinh x\cos\t\g_{1}\hat\g
  +\sinh x\sin\t\cos\vp\g_{2}\hat\g + \sinh x\sin\t\sin\vp\g_{3}\hat\g)\no
  &\cdot [-\m_{1} \g_{47}\hat\g
  -\m_{2} e^{-\xi_{1}\g_{47}}e^{-\xi_{2}\g_{58}}\g_{57}\hat\g
  -\m_{3} e^{-\xi_{1}\g_{47}}e^{-\xi_{3}\g_{69}}\g_{67}\hat\g
  +\m_{4} e^{-\xi_{1}\g_{47}}e^{-\xi_{4}\g_{10}\hat\g}\g_{7}]\no
  \g_{8}\CM&=\CM(\cosh xe^{t\g_{0}\hat\g} + \sinh x\cos\t\g_{1}\hat\g
  +\sinh x\sin\t\cos\vp\g_{2}\hat\g + \sinh x\sin\t\sin\vp\g_{3}\hat\g)\no
  & \cdot [-\m_{1} e^{-\xi_{1}\g_{47}}e^{-\xi_{2}\g_{58}}\g_{48}\hat\g
  -\m_{2} \g_{58}\hat\g
  -\m_{3} e^{-\xi_{2}\g_{58}}e^{-\xi_{3}\g_{69}}\g_{68}\hat\g
  +\m_{4} e^{-\xi_{2}\g_{58}}e^{-\xi_{4}\g_{10}\hat\g}\g_{8}]\no
  \g_{9}\CM&=\CM(\cosh xe^{t\g_{0}\hat\g} + \sinh x\cos\t\g_{1}\hat\g
  +\sinh x\sin\t\cos\vp\g_{2}\hat\g + \sinh x\sin\t\sin\vp\g_{3}\hat\g)\no
  &\cdot [-\m_{1} e^{-\xi_{1}\g_{47}}e^{-\xi_{3}\g_{69}}\g_{49}\hat\g
  -\m_{2}e^{-\xi_{2}\g_{58}}e^{-\xi_{3}\g_{69}}\g_{59}\hat\g
  -\m_{3} \g_{69}\hat\g
  +\m_{4} e^{-\xi_{3}\g_{69}}e^{-\xi_{4}\g_{10}\hat\g}\g_{9}]\no
  \g_{10}\CM&=\CM(\cosh xe^{t\g_{0}\hat\g} + \sinh x\cos\t\g_{1}\hat\g
  +\sinh x\sin\t\cos\vp\g_{2}\hat\g + \sinh x\sin\t\sin\vp\g_{3}\hat\g)\no
  &\cdot [-\m_{1} e^{-\xi_{1}\g_{47}}e^{-\xi_{4}\g_{10}\hat\g}\g_{410}\hat\g
  -\m_{2}e^{-\xi_{2}\g_{58}}e^{-\xi_{4}\g_{10}\hat\g}\g_{510}\hat\g
  -\m_{3} e^{-\xi_{3}\g_{69}}e^{-\xi_{4}\g_{10}\hat\g}\g_{610}\hat\g
  +\m_{4}\g_{10}]\no
  \g_y\CM&=\CM(\cosh xe^{t\g_{0}\hat\g} + \sinh x\cos\t\g_{1}\hat\g
  +\sinh x\sin\t\cos\vp\g_{2}\hat\g + \sinh x\sin\t\sin\vp\g_{3}\hat\g)\no
  & \cdot
  [-\m_{1}\m_{2}e^{-\xi_{1}\g_{47}}e^{-\xi_{2}\g_{58}}(\g_{48}+\g_{57})\hat\g
  -\m_{1}\m_{3}e^{-\xi_{1}\g_{47}}e^{-\xi_{3}\g_{69}}(\g_{49}+\g_{67})\hat\g\no
  &
  \quad+\m_{1}\m_{4}e^{-\xi_{1}\g_{47}}e^{-\xi_{4}\g_{10}\hat\g}(\g_{7}-\g_{410}
  \hat\g)-
  \m_{2}\m_{3}e^{-\xi_{2}\g_{58}}e^{-\xi_{3}\g_{69}}(\g_{68}+\g_{59})
  \hat\g\no
  &\quad~+ \m_{2}\m_{4}e^{-\xi_{2}\g_{58}}e^{-\xi_{4}\g_{10}\hat\g}
  (\g_{8}-\g_{510}\hat\g) + \m_{3}\m_{4}e^{-\xi_{3}\g_{69}}e^{-\xi_{4}
  \g_{10}\hat\g}(\g_{9}-\g_{610}\hat\g)\no
  &\quad~~-(\m_{1}^{2}\g_{47}+\m_{2}^{2}\g_{58}+\m_{3}^{2}\g_{69}
  + \m_{4}^{2}\g_{10}\hat\g)\hat\g]
\end{align}

\begin{align}
  e^{t\g_0\hat\g}\g_0e^{t\g_0\hat\g}=\g_0
\end{align}


\end{document}